\documentstyle[11pt]{book}
\input amssym.def
\input amssym

\def\R{{\Bbb R}}

\def\C{{\Bbb C}}

\newcommand{\be}{\begin{equation}}
\newcommand{\ee}{\end{equation}}

\def\bea{\begin{eqnarray}}
\def\eea{\end{eqnarray}}

\def\cD{{\cal D}} \def\cE{{\cal E}} \def\cF{{\cal F}}
\def\cN{{\cal N}}

\newcommand{\eq}[1]{(\ref{#1})}

\begin{document}
\chapter*{ }
\vskip -5truecm

\centerline {\Large\bf Quantization and eigenvalue distribution} 
\centerline {\Large\bf of noncommutative scalar field theory}
\vskip 10truemm
\centerline {\bf Harold Steinacker}  
\vskip 5truemm
\centerline{Institute for Theoretical Physics, University of Vienna}
\centerline{Boltzmanngasse 5, A-1090 Vienna, Austria}

{\renewcommand{\thefootnote}{}
}

\setcounter{page}{1}

\markboth{ } 
{ }  

\vskip 5truemm

\begin{quote}
\small{\bf Abstract.} {\it

The quantization of noncommutative scalar field theory 
is studied from the matrix model point of view, exhibiting
the significance of the eigenvalue distribution. 
This provides a new framework to study 
renormalization, and predicts a
phase transition in the noncommutative $\phi^4$ model.  
In 4-dimensions, the corresponding 
critical line is found to terminate at a non-trivial point.  
 }
\end{quote}

\begin{quote}
\small{\bf Key words:} {\it Noncommutative field theory, phase transition, 
matrix model, renormalization }
\end{quote}
\vskip 5truemm

\centerline {\sc 1. INTRODUCTION: NC FIELD THEORY AND UV/IR MIXING}
\vskip 3truemm

Noncommutative (NC) field theory has been studied intensively in recent years, 
see e.g. \cite{douglas, Szabo} and references therein. 
Part of the motivation has been the hope that the UV divergences of Quantum Field
Theory should be controlled and perhaps regularized by noncommutativity,
which in turn should be related to quantum fluctuations of geometry 
and quantum gravity.
In the simplest case of the quantum plane 
$\R^{d}_\theta$, the coordinate functions $x_i, i=1,...,d$ satisfiy
the canonical commutation relations   
\be
[x_i,x_j] = i \theta_{ij}.
\label {CC}
\ee
We assume for simplicity that $\theta_{ij}$ is 
nondegenerate and maximally-symmetric, characterized by a single
parameter $\theta$ with dimension $({\rm length})^2$. This
implies uncertainty relations among the coordinates,
\be
\Delta x_i \Delta x_j \geq \frac 12 \theta_{ij} \approx \theta.
\label{uncertainty}
\ee
This leads to an energy scale
\be
\Lambda_{NC} = \sqrt{\frac 1{\theta}},
\ee
and noncommutativity is expected to be important for energies
beyond $\Lambda_{NC}$.

The most naive guess would be that $\Lambda_{NC}$ plays the role of a UV cutoff
in Quantum Field Theory.
However, this turns out not to be the case. 
The quantization of NC field theory 
not only suffers from divergences in the UV, but also from new divergencies
in the IR, which are in general not under control up to now.  
This type of behavior can to some extent be expected from \eq{CC}.
To understand this, expand a scalar field on $\R^{d}_\theta$ 
in a basis of plane waves,
$\phi(x) = \int d^d p \,\phi(p) \exp(i p x)$, and
consider the range of momenta $p$. Formally, $p \in \R^d$
has neither an IR nor a UV cutoff. 
As usual in field theory, it is necessary however
to impose cutoffs $\Lambda_{IR} \leq p \leq \Lambda_{UV}$, with the hope that the 
dependence on the cutoffs can be removed in the end;  
this is essentially the defining property of a renormalizable field theory.

In view of \eq{uncertainty}, if $\Delta x_i$ is minimized according to 
$\Delta x_i = \Lambda_{UV}^{-1}$, 
it implies that some $\Delta x_j \geq \theta \Lambda_{UV}$. 
This means that momenta near
$\Lambda_{UV}$ are inseparably linked with momenta in the IR,
and suggests $\Lambda_{UV} \Lambda_{IR} = \Lambda_{NC}^2$.
A consistent way to impose IR and UV cutoffs in NC field theory is to
replace $\R^n_\theta$ by some compact NC spaces such as fuzzy tori, fuzzy 
spheres etc. The algebra of functions is then given by a
finite matrix algebra $Mat(\cN\times\cN,\C)$. One then finds 
quite generally 
\be
\Lambda_{IR} =  \sqrt{\frac 1{N\theta}} \quad\ll\quad \Lambda_{NC} =
\sqrt{\frac 1{\theta}} \quad\ll\quad \Lambda_{UV} = \sqrt{\frac N{\theta}}
\ee
 where $\cN \approx N^{d/2}$ 
(depending on the details of the compact NC space) is the 
dimension of the representation space. This shows that
the scales in the UV and IR are intimately related, which is 
the origin of $UV/IR$ mixing in NC field theory. It is then 
not too surprising that a straightforward Wilsonian renormalization
scheme involving only $\Lambda_{UV}$ 
will generically fail \cite{SMV}.

The remainder of this paper is a qualitative discussion of a 
new approach to quantization of euclidean NC scalar field theory,
which was proposed in \cite{NCscalarFT} with the goal to 
take into account more properly the specific features of NC field theory.

\vskip 5truemm
\centerline {\sc 2. NONCOMMUTATIVE SCALAR FIELD THEORY}
\vskip 3truemm

Consider the scalar $\phi^4$ model on a suitable $d$-dimensional NC space,
\be
S =  \int d^d x\, \left(\frac 12  \phi \Delta \phi 
+ \frac 12 m^2 \phi^2 + \frac{\lambda}{4} \phi^4\right)
 = S_{kin} + \int d^d x\, V(\phi).
\label{Rd-action}
\ee
We assume that the algebra of functions on the NC space is represented
as operator algebra on some {\em finite-dimensional} Hilbert space 
(this is the defining property of the so-called fuzzy spaces), so that
$\phi \in Mat(\cN\times \cN,\C)$ is a Hermitian matrix.
The integral is then replaced by the 
appropriately normalized trace,  $\int d^d x = (2\pi\theta)^{n/2} Tr()$.

In this finite setup, the quantization is defined by an integral over all
Hermitian matrices, e.g.
\bea
Z &=& \int [\cD \phi] e^{-S},  \label{Z-full}\\
\langle \phi_{i_1 j_1} \cdots \phi_{i_n j_n} \rangle &=& 
\frac 1Z \int [\cD \phi] e^{-S}\;  \phi_{i_1 j_1} \cdots \phi_{i_n j_n}
\eea
As a warm-up, we recall the analogous but simpler case of 

\paragraph{Pure matrix models.}

These are defined by an action of the form
\be
S^{M.M.} = \cN Tr(V(\phi))
\ee
for some polynomial $V(\phi)$, 
where again $\phi \in Mat(\cN\times \cN,\C)$ is a Hermitian matrix.
These pure matrix models can be solved exactly by the following change 
of coordinates in matrix space:
\be
\phi = U^{-1} D U
\ee
where $D$ is a diagonal matrix with real eigenvalues $\phi_i$, and $U \in U(\cN)$.
Using $[\cD \phi] = \Delta^2(\phi_i) dU d\phi_1 ... d \phi_N$, where $dU$ denotes the 
Haar measure for $U(\cN)$ and $\Delta(\phi_i) = \prod_{i<j}(\phi_i-\phi_j)$
is the Vandermonde-determinant.
The integral over $dU$ can be carried out trivially,
reducing the path integral to an integral over eigenvalues,
\be
Z =  \int d\phi_i \Delta^2(\phi_i-\phi_j)\exp(-\cN \sum_i V(\phi_i)) 
= \int d\phi_i \exp(-S_{eff}^{M.M}(\vec\phi)) 
\label{Z-MM}
\ee
with an ``effective action'' of the form
\be
\exp(-S_{eff}^{M.M.}(\phi_i)) = \exp(\sum_{i\neq j}
\log|\phi_i-\phi_j| -\cN \sum_i V(\phi_i)).
\label{Seff-MM}
\ee
We can assume that the eigenvalues are ordered,
$\phi_1 \leq \phi_2 \leq ... \leq \phi_\cN$, and
denote with $\cE_\cN$ 
the space of ordered $\cN$-tupels $\vec \phi = (\phi_1, ..., \phi_\cN)$. 
They will be interpreted as 
coordinates of a point in the space of eigenvalues $\cE_\cN$.

The remaining integral over $\phi_i$ can be evaluated with a variety of techniques 
 such as orthogonal polynomials, Dyson- Schwinger resp. loop equations, 
or the saddle-point method. It turns out that the large $\cN$ limit
is correctly reproduced by the saddle-point method,
because the Vandermone-determinant 
$\Delta^2(\phi_i) = \exp(\sum_{i\neq j} \ln|\phi_i-\phi_j|)$ corresponds to a 
strongly repulsive potential between the eigenvalues, which
leads to a strong localization in $\cE_\cN$.
This means that the effective action $\exp(-S_{eff}^{M.M.}(\vec\phi))$ 
is essentially a delta-function in $\cE_\cN$,
\be
\exp(-S_{eff}^{M.M.}(\vec \phi)) \approx \exp(-\cN Tr(V(\vec \phi_0))) \,
 \delta(\vec \phi - \vec \phi_0).
\label{localiz-delta}
\ee
One now proceeds to determine the 
localization of the maximum $\vec \phi_0 \in \cE_\cN$ by solving 
$\frac{\partial}{\partial \phi_i} S_{eff} =0$, which can be written as an 
integral equation in the large $\cN$ limit. 
Expectation values can then be computed as 
\be
\langle f(\vec\phi)\rangle =  f(\vec\phi_0) 
= f(\langle\vec\phi \rangle)
\ee
which is indeed correct in the large $\cN$ limit of pure matrix models, 
consistent with  \eq{localiz-delta}.
This  "factorization'' of 
expectation values is characteristic for a delta-function 
integral density, and can be used to test the localization hypothesis
\eq{localiz-delta} resp. 
the sharpness of the (approximate) delta function.

\paragraph{Strategy for the full model.}

This analogy suggests to apply similar techniques
also for the field theory \eq{Z-full}. 
At first sight, this may appear impossible, 
because the action is no longer invariant under
$U(\cN)$. However, let us assume that we could somehow evaluate the 
``angular integrals''
\be
\int dU \exp(-S_{kin}(U^{-1}(\phi) U))  = : e^{-\cF(\vec\phi)},
\ee
Note that the resulting $\cF(\phi_i)$ is a totally symmetric, analytic function 
of the eigenvalues $\phi_i$ of the 
field $\phi$, due to the integration over the unitary matrices $U$.
Using this definition, the partition function 
can indeed be cast into the same form as \eq{Z-MM},
\be
Z  = \int d \phi_i \Delta^2(\phi_i)\exp(-\cF(\vec\phi) -(2\pi\theta)^{d/2}\;Tr V(\vec\phi))
=  \int d\phi_i \exp(-S_{eff}(\vec\phi)) 
\label{Z-F}
\ee
(with the obvious interpretation of $Tr V(\vec\phi)$), defining 
\be
\exp(-S_{eff}(\vec\phi)) =\Delta^2(\phi_i)\exp(- \cF(\vec\phi) - (2\pi\theta)^{d/2}\;Tr V(\vec\phi)).
\label{S-eff}
\ee
Of course we are only allowed
to determine observables depending on the eigenvalues with this effective action
(the extension to other observables is discussed in \cite{NCscalarFT}).
These are determined by 
$S_{eff}(\vec\phi)$ in the same way as for pure matrix models, 
since  the degrees of freedom
related to $U$ are integrated out.  
This demonstrates that all thermodynamic properties
of the quantum field theory, in particular
the phase transitions,  are determined
by $S_{eff}(\vec\phi)$  and  the resulting eigenvalue distribution 
in the large $\cN$ limit. 
The advantage of this
formulation is that it is very well suited to include interactions, 
and naturally extends to the non-perturbative domain. 

So far, all this is exact but difficult to apply, since we will not be
able to evaluate $\cF(\vec\phi)$ exactly. 
However, the crucial point in \eq{S-eff} is the presence of the 
Vandermonde-determinant 
$\Delta^2(\phi_i) = \exp(\sum_{i\neq j}\ln|\phi_i-\phi_j|)$,  
which prevents the eigenvalues from 
coinciding and strongly localized them. This effect cannot be 
canceled by $\cF(\vec\phi)$, because it ts analytic. 
This and the analogous exact analysis of the pure matrix models
suggests the following {\em central hypothesis} of our approach:
\eq{S-eff} is strongly localized, and
the essential features of the full QFT
are reproduced by the following approximation
\be
\exp(-S_{eff}(\vec\phi)) \approx \exp(-\cF(\vec\phi_0) 
- (2\pi\theta)^{d/2}\;Tr V(\vec\phi_0)) \,
\delta(\vec \phi - \vec \phi_0),
\label{localiz-delta-full}
\ee
where $\vec \phi_0$ denotes the maximum (saddle-point) of $S_{eff}(\vec\phi)$.
The 
remaining integral over the eigenvalues $\phi_i$ can 
hence be evaluated by the saddle-point method.
This will be justified to some extent and made more precise
below, by studying
$\cF(\vec\phi)$ using the free case. 
If $\cF(\vec\phi)$ is
known, the free energy can be found by determining the minimum
$\phi_0$ of $S_{eff}(\vec\phi)$.
However, \eq{localiz-delta-full} strongly suggests that also
correlation functions can be computed by restricting 
the full matrix integral to the
dominant eigenvalue distribution  $\vec\phi_0$. This 
would provide nonperturbative control over the full NC QFT.

It is important to keep in mind that
a saddle-point approximation {\em before} integrating out $U(\cN)$ would be
complete nonsense. It is only appropriate for the effective action 
\eq{localiz-delta-full}.
The possibility of separating the path integral into
these 2 steps is only available for NC field theory.

\vskip 5truemm
\centerline {\sc 3. THE CASE OF FREE FIELDS}
\vskip 3truemm

For free fields, we can verify the validity of the basic hypothesis
\eq{localiz-delta-full}, and compute the unknown function 
$\cF(\vec\phi)$ explicitly at least in some domain. This 
will then be applied in the interacting case.

To check localization in the sense of \eq{localiz-delta-full}, 
we need to show that
the expectation values of all observables which depend only on the eigenvalues
$\vec\phi$ factorize, and can be computed by evaluating the observable at a 
specific point $\vec  \phi_0$. 
A complete set of such observables is given by the product of traces of 
various powers of $\phi$, which in field theory language is 
$\langle(\int d^d x \phi(x)^{2n_1})...(\int d^d x \phi(x)^{2n_k}) \rangle$.
It is shown in \cite{NCscalarFT} using Wicks theorem
that  in the large $N$ resp. $\Lambda_{UV}$ limit,
\bea
\frac{\langle\big( \frac 1V\int d^d x \phi(x)^{2n_1}\big)
\big( \frac 1V\int d^d x \phi(x)^{2n_k}\big) \rangle}
{\langle  \frac 1V\int d^d x \phi(x)^{2} \rangle^{n_1}
\langle  \frac 1V\int d^d x \phi(x)^{2} \rangle^{n_k}}
&=& \frac{\langle \frac 1V\int d^d x \phi(x)^{2n_1}\rangle}
  {\langle  \frac 1V\int d^d x \phi(x)^{2} \rangle^{n_1}}\; 
\;\frac{\langle \frac 1V\int d^d x \phi(x)^{2n_k} \rangle}
{\langle  \frac 1V\int d^d x \phi(x)^{2} \rangle^{n_k}} \nonumber\\
&=& N_{Planar}(2n_1)\; ... \;N_{Planar}(2n_k).
\label{cluster}
\eea
Here $N_{Planar}(2n)$ is the number of planar contractions of a
vertex with $2n$ legs. 
The non-planar contributions always involve oscillatory
integrals, and do not contribute to the above ratio in the large $N$ limit. 

This is exactly the desired factorization property, which implies
that the effective action of the eigenvalue sector localizes 
as in \eq{localiz-delta-full}, after suitable rescaling.
This is only true for 
noncommutative field theories, where nonplanar (completely contracted) diagrams
are suppressed by their oscillating behavior, so that only
the planar diagrams contribute\footnote{this is only true for observables of
the above type, not e.g. for propagators}.

To complete the analysis of the free case, we need to find the dominant
eigenvalue distribution $\vec\phi_0$. That is an easy task using the known 
techniques from pure matrix models described above. 
Writing 
$\varphi(s) = (\phi_0)_j, \quad   s = \frac j{\cN}$ for $s \in [0,1]$ 
in the continuum limit, 
the saddle-point $\vec\phi_0$ then corresponds to a density of eigenvalues 
$\rho(\varphi) = \frac{ds}{d\varphi}$ 
with $\int_{-\infty}^{\infty} \rho(p) dp =1$. This turns out to be
the famous Wigner semi-circle law 
\be
\rho(p) = \left\{\begin{array}{ll}\frac 2{\pi} \sqrt{1-p^2} & \;\; p^2<1\\
                           0, & \mbox{otherwise}.
                \end{array}\right.
\label{wigner-law}
\ee
This is the same as for the pure Gaussian matrix model, and it follows that
the effective action for the eigenvalue sector is given by 
\be 
S_{eff}^{free}(\vec\phi)  
= f_0(m) -\sum_{i\neq j}\ln|\phi_i-\phi_j| + \frac{2\cN}{\alpha_0^2(m)} (\sum\phi_i^2).
\label{eff-act-eig}
\ee
Here
$f_0(m)$ is some numerical function of $m$ which is not important here, 
and 
\be
\alpha_0^2(m) = \left\{\begin{array}{ll} \frac{1}{4\pi^{2}}\;
\Lambda_{UV}^2 \left(1 - \frac{m^2}{\Lambda_{UV}^2}\ln(1+\frac{\Lambda_{UV}^2}{m^2})\right), & d=4\\
                 \frac{1}{\pi}\ln(1+\frac{\Lambda_{UV}^2}{m^2})  , & d=2
                \end{array}\right.
\label{alpha}
\ee 
depending on the dimension of the field theory.

\vskip 5truemm
\centerline {\sc 3. APPLICATION: THE $\phi^4$ MODEL}
\vskip 3truemm

We can now apply the result  \eq{eff-act-eig} to the interacting case.
Consider for example the $\phi^4$ model \eq{Rd-action},
which is obtained from the free case by adding 
a potential of the form
$$
S_{int}(\phi) = \frac{\lambda}{4}\int d^d x \phi^4(x) 
= \frac{\lambda}{4} (2\pi\theta)^{d/2}
 Tr \phi^4,
$$
The goal is to derive some properties (in particular
thermodynamical, but also others) of the interacting model.
The basic hypothesis is again \eq{localiz-delta-full}, where the 
eigenvalue distribution $\vec\phi_0$ in the interacting case
is to be determined. 

It is quite easy to understand the main effect of the interaction term
using the above results \eq{eff-act-eig}: 
since $S_{int}$ only depends on the eigenvalues, it seems
very natural to simply add $S_{int}$ to \eq{eff-act-eig}, suggesting
\be
Z_{int} = \int d\phi_i e^{-S_{eff}^{free}(\vec\phi) - S_{int}(\vec\phi)}.
\label{Z-perturb}
\ee
To justify this, we can expand the
interaction term into a power series in $\lambda$. 
Using the factorization property \eq{cluster}, 
we can immediately compute any expectation value of 
observables which depend only on the eigenvalues, and sum up the 
expansion in $\lambda$. As in the case of pure matrix models \cite{BIPZ},
the effect of this perturbative computation is reproduced
for $\cN \to \infty$ 
by the second (nonperturbative) point of view, which is that
\eq{Z-perturb} is localized at a new eigenvalue distribution
$\vec\phi_0^{\lambda}$, which now depends on the interaction.
It can again be found by the saddle-point approximation.

We conclude with some remarks and results of this analysis \cite{NCscalarFT}:

\begin{enumerate}
\item
A caveat: the validity of \eq{eff-act-eig} has been established only 
``locally'', i.e. the function $\cF(\vec\phi)$ is really known only
for $\vec \phi \approx\vec \phi_0$. The only free parameter available 
is the mass\footnote{this can be enhanced by adding other
  terms such as the one in \cite{wulki}, which will be pursued
  elsewhere}, 
which enters in \eq{eff-act-eig} through $\alpha_0$ and allows
to test a one-dimensional submanifold of $\cE_\cN$. Therefore we 
can trust \eq{eff-act-eig} only if $\vec\phi_0^{\lambda} \approx \vec\phi_0$.
This leads  to the second observation:
\item
The physical properties of the interacting model will be close to those of 
the free case only if $\vec\phi_0^{\lambda} \approx \vec\phi_0$.
In particular, the correlation length (i.e. physical mass) of the interacting model
will be reproduced best by the free action whose EV distribution 
$\vec\phi_0$ is closest to $\vec\phi_0^{\lambda}$.
Working this out \cite{NCscalarFT} leads to a very simple understanding of mass
renormalization. Namely, it turns out that the bare mass must be
adjusted to
\be
m^2  =  m_{phys}^2 - \frac{3}{16\pi^{2}}\;
   \Lambda_{UV}^2 \left(1 -\; \frac{m_{phys}^2}{\Lambda_{UV}^2}\;
\ln(\frac{\Lambda_{UV}^2}{m_{phys}^2})\right)\; \lambda.
\label{mass-ren-4d}
\ee
in the 4-dimensional case. This reproduces precisely the result of 
a conventional one-loop computation, however it is now based on
a nonperturbative analysis and not just a formal expansion.
\item
Stretching somewhat the range where \eq{eff-act-eig} has been tested, 
one can study the
thermodynamical properties and phase transitions of the interacting model. 
Indeed a phase transition is found at the point where the eigenvalue distribution
$\vec\phi_0^\lambda$ breaks up into 2 disjoint pieces ("2 cuts"), due to the
interaction term. 
This is expected to be 
the transition between the ``striped'' and disordered phase 
found in NC scalar field theory \cite{gubser,bieten}.
In the most interesting case of 4 dimension, a critical line 
is found which ends at a nontrivial critical point $\lambda_c >0$ 
\cite{NCscalarFT}. 
This is strongly
suggestive for a {\em nontrivial} NC $\phi^4$ model in 4 dimensions,
since the endpoint of the critical line $\lambda_c >0$ should correspond to 
a fixed-point under a suitable RG flow.

Since the mechanism of the phase transition is very generic and does
not depend on the details of the potential, the 
qualitative features of this result are expected to be correct.

\end{enumerate}
Full details can be found in \cite{NCscalarFT}.

\paragraph{Acknowledgments} It is a pleasure to thank the organizers
of the II. Southeastern European Workshop BW 2005, Vrnjacka Banja,
Serbia
and the XIV.  Workshop of Geometric Methods in Physics,
Bialowieza, Poland 
for providing an enjoyable and stimulating
atmosphere. This work was supported by the FWF project
P16779-N02.

\nopagebreak
\renewcommand {\bibname} {\normalsize \sc References}
\nopagebreak

{}


\begin{thebibliography}{99}

\bibitem{douglas} 
M.~R.~Douglas and N.~A.~Nekrasov,
``Noncommutative field theory,''
{\it Rev.Mod.Phys.  } {\bf 73} (2001) 977; {\tt [hep-th/0106048]}

\bibitem{Szabo}
R. Szabo, ``Quantum Field Theory on Noncommutative Spaces''.
  {\it Phys.Rept.} {\bf 378} (2003) 207-299; {\tt [hep-th/0109162]}

\bibitem{SMV}  S. Minwalla, M. Van Raamsdonk, N. Seiberg
``Noncommutative Perturbative Dynamics''.
{\em  JHEP}  {\bf  0002} (2000) 020; {\tt [hep-th/9912072]}

\bibitem{NCscalarFT} H. Steinacker, 
"A non-perturbative approach to non-commutative scalar field theory".
{\it JHEP}{\bf 0503} (2005) 075; {\tt [hep-th/0501174]}

\bibitem{wulki}  H. Grosse, R. Wulkenhaar, 
``Renormalisation of $\phi^4$-theory on noncommutative $R^4$ in the matrix
base''.  {\tt [hep-th/0401128]}

\bibitem{BIPZ}
E.~Brezin, C.~Itzykson, G.~Parisi and J.~B.~Zuber,
``Planar Diagrams,''
 {\it  Commun.\ Math.\ Phys.}  {\bf 59} (1978) 35.

\bibitem{gubser} S. Gubser, S. Sondhi, 
"Phase structure of non-commutative scalar field theories''.
{\it Nucl.Phys.}{\bf B605} (2001) 395-424; {\tt [hep-th/0006119]}


\bibitem{bieten}  W. Bietenholz, F. Hofheinz, J. Nishimura, 
``Phase diagram and dispersion relation of the non-commutative $\lambda
\phi^{4}$ model in d=3''.  {\it JHEP } {\bf 0406} (2004) 042; 
 X. Martin, "A matrix phase for the $\phi^4$ scalar
  field on the fuzzy sphere''.  {\em JHEP} {\bf 0404} (2004) 077




\end{thebibliography}
\end{document}